\documentclass[twocolumn,groupedaddress,amsmath,amssymb,prl,aps]{revtex4}

\usepackage{graphicx}
\usepackage{dcolumn}
\usepackage{bm}
\usepackage{soul} 
\usepackage{xcolor} 
\usepackage{amsmath} 
\usepackage{lipsum}

\begin{document}

\title{Theoretical approach to ferroelectricity in hafnia and related
  materials}

\author{Hugo Aramberri$^{1}$ and Jorge \'I\~niguez$^{1,2}$}

\affiliation{
  \mbox{$^{1}$Materials Research and Technology Department,
    Luxembourg Institute of Science and Technology (LIST),} \mbox{Avenue
    des Hauts-Fourneaux 5, L-4362 Esch/Alzette,
    Luxembourg}\\
 \mbox{$^{2}$Department of Physics and Materials Science, University
   of Luxembourg, Rue du Brill 41, L-4422 Belvaux, Luxembourg}}

\maketitle

{\bf Hafnia ferroelectrics combine technological promise and
  unprecedented behaviors. Their peculiarity stems from the many
  active extrinsic mechanisms that contribute to their properties and
  from a continuously growing number of novel intrinsic
  features. Partly because of their unconventional nature, basic
  questions about these materials remain open and one
    may wonder about the pertinence of some frequent theoretical
    assumptions. Aided by first-principles simulations, here we show
  that, by adopting an original high-symmetry
    reference phase as the starting point of the analysis, we can
    develop a mathematically simple and physically transparent
    treatment of the ferroelectric state of hafnia. The proposed
  approach describes hafnia as a uniaxial ferroic, as
    suggested by recent studies of (woken-up) samples with well
    developed polarization. Also, it is compatible with the occurrence
    of polar soft modes and proper ferroelectric order.  Further, our
  theory provides a straightforward and unified description of all
  low-energy polymorphs, shedding light into old questions (e.g., the
  prevalence of the monoclinic ground state), pointing at exciting
  possibilities (e.g., an antiferroelastic behavior) and facilitating
  the future development of perturbative theories (from Landau to
  second-principles potentials). Our work thus yields a deeper
  understanding of hafnia ferroelectrics, improving our ability to
  optimize their properties and induce new ones.}

Hafnia ferroelectrics~\cite{boscke11,muller12} -- including HfO$_{2}$,
Hf$_{1-x}$Zr$_{x}$O$_{2}$, ZrO$_{2}$ and doped variations -- attract
attention because of their technological promise \cite{bohr07} and
surprising properties, from their resilient polar order in
nanostructures \cite{boscke11} to their tunable
piezoresponse~\cite{dutta21}. Understanding their behavior is
challenging, though, a major difficulty coming from the variety of
intrinsic and extrinsic factors that influence the observed
properties. Even if we focus on the intrinsic features of perfect
crystals, as those considered in quantum computer simulations, these
materials prove exceedingly intriguing.

Recent works show that, in the case of the most
common ferroelectric phase of hafnia (orthorhombic with space group
$Pca2_{1}$), some most basic questions -- e.g., what
  should be the centrosymmetric state that can be used as a reference
  to compute the polarization \cite{choe21,qi22} -- are not resolved
yet. In fact, we contend that the common
 theoretical treatments
  \cite{reyes-lillo14,delodovici21} -- which assume the well-known
  tetragonal ($P4_{2}/nmc$) or cubic ($Fm\bar{3}m$) phases as
  reference structure -- are not ideally suited to discuss the
  behavior of the woken-up ferroelectric phase of these
  materials. Indeed, such approaches yield involved models that hamper
  the discussion of the relevant low-energy landscape. Then, based on
first-principles simulations, we show that there exists an alternative
reference state that solves these issues while
  providing us with an appealing picture of the ferroelectric state
  and its properties.

\begin{figure}
    \centering
    \includegraphics[width=\columnwidth]{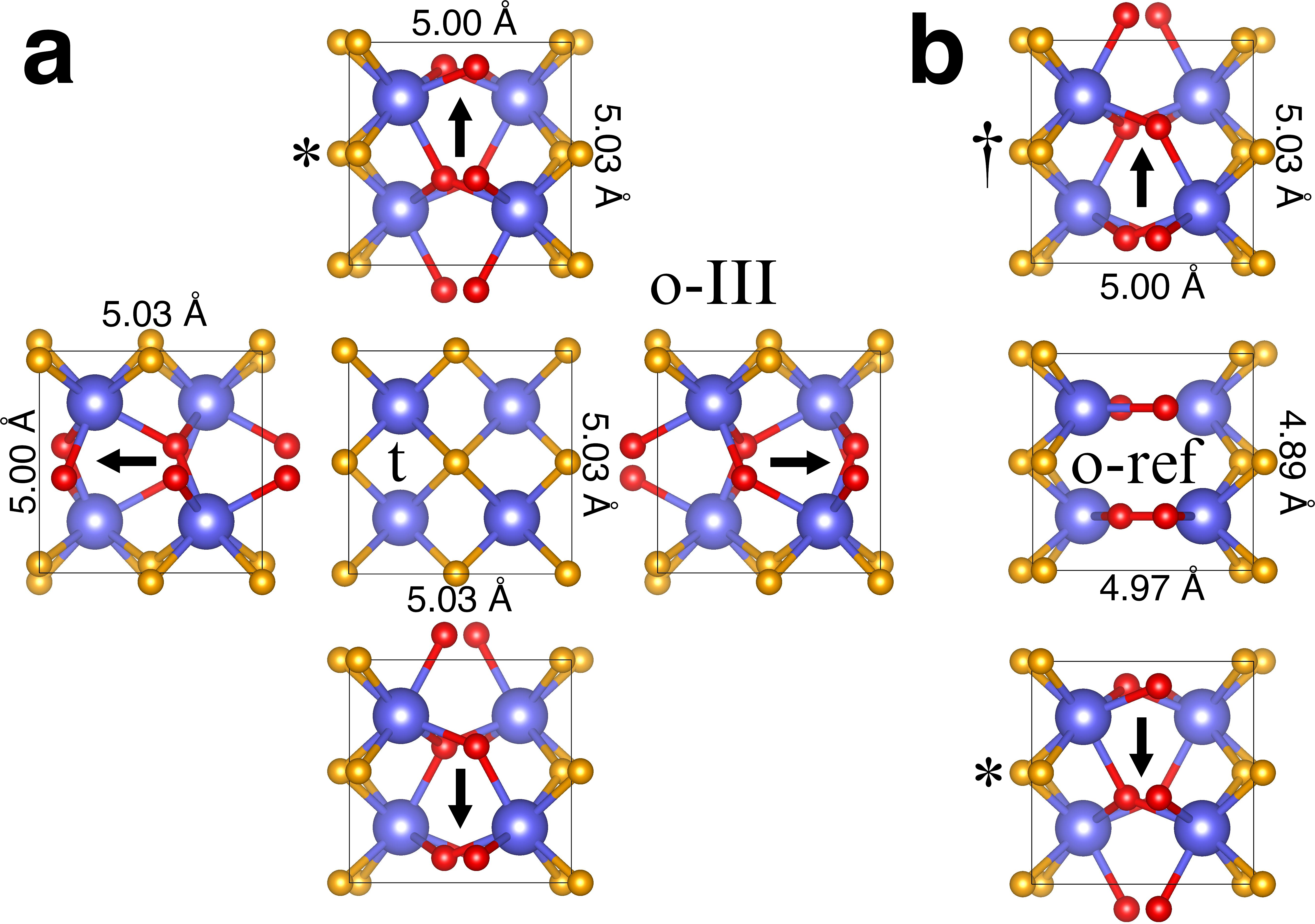}
    \caption{{\bf Ferroelectric domains expected in hafnia.}  {\bf a}
      shows the tetragonal (t) phase of hafnia (center) and the four
      orthorhombic ferroelectric (o-III) variants it leads to. {\bf b}
      shows the orthorhombic centrosymmetric phase (o-ref) we propose
      as reference (center) and the two o-III domains it leads to. The
      active oxygens, responsible for the development of the
      spontaneous polarization, are shown in red; the other oxygens
      are shown in orange. Black arrows indicate the spontaneous
      polarization, which goes against the displacement of the active
      oxygens from the reference structure. The computed polarization
      with respect to the t-phase for the structure marked with an
      asterisk in {\bf a} is $+0.54$~C~m$^{-2}$; by contrast, the
      polarization with respect to the o-ref phase of the structure
      marked with an asterisk in {\bf b} is $-0.68$~C~m$^{-2}$, while
      the one marked with a dagger presents $+0.68$~C~m$^{-2}$. (As
      shown in Supplementary Note~1, $+0.54$~C~m$^{-2}$ and
      $-0.68$~C~m$^{-2}$ differ by two polarization quanta.) Lattice
      constants are indicated.}
    \label{fig:domains}
\end{figure}

Let us first discuss the issues we see in current
treatments. Figure~\ref{fig:domains}{\bf a} sketches the
tetragonal phase most frequently considered as the
  centrosymmetric reference in discussions of ferroelectricity in
  hafnia ($P4_{2}/nmc$ space group, ``t'' in the following). The
  figure also shows the variants of the $Pca2_{1}$ ferroelectric
phase (``o-III'' in the following) obtained from this t-reference. The
distortion from t to o-III breaks the tetragonal 4-fold axis, and the
resulting polarization that can adopt four symmetry-equivalent
orientations within the plane of the figure, $(\pm P,0)$ and $(0,\pm
P)$. Accordingly, one would expect to find four different domains in
hafnia samples. For example, we could have 90$^{\circ}$ ferroelectric
and ferroelastic domain walls separating regions with ${\bf P}=(P,0)$
and ${\bf P}=(0,P)$. Such 90$^{\circ}$ boundaries have
  indeed been observed at intermediate stages during the electric
  cycling frequently required to ``wake up'' hafnia samples
  \cite{shimizu18,lederer21}, as the material visits intermediate
  states with coexisting regions of in-plane (as-grown) and
  out-of-plane (poled) polarization. Similar 90$^{\circ}$ walls have
also been observed upon ferroelectric switching in some
samples~\cite{zhou22a}. The reported 90$^{\circ}$
  walls would be nominally charged and of very high energy according
  to Density Functional Theory (DFT) calculations~\cite{ding20}, which
  suggests yet-undetermined screening mechanisms might be at play. At
  any rate, beyond some lingering mysteries, the observations clearly
  indicate that in some situations -- particularly during wake-up
  cycling -- hafnia behaves as a ferroelastic biaxial (even triaxial)
  material, which demands a theory based on a tetragonal (even cubic)
  high-symmetry reference structure.

By contrast, we think there are reasons to believe
  that the treatment of woken-up hafnia samples could be much
  simpler. Experimental studies \cite{li23} suggest that woken-up
  hafnia and zirconia samples often present a coexistence of phases,
  including the o-III ferroelectric state, the well-known monoclinic
  ground state (m-phase) and other low-energy orthorhombic polymorphs
  (e.g., those usually denoted o-I and
  o-I*~\cite{azevedoantunes22}). Remarkably, microscopy images suggest
  that such polymorphs are separated by boundaries of essentially zero
  width~\cite{du21,li23}. In fact, such polytypic states appear to be
  a well oriented structure where all the polymorphs share a common
  (anti)polar axis of sorts. (This statement will be made precise
  below.) These conclusions are reinforced by recent first-principles
  investigations of ferroelectric switching in hafnia~\cite{silva23}:
  the most likely paths involve intermediate structures that share
  atomic motifs with the low-energy phases of the material (m, o-I and
  o-I*), all displaying a common (anti)polar axis. Hence, a picture of
  woken-up hafnia as a uniaxial ferroelectric starts to emerge.

\begin{figure}
    \centering
    \includegraphics[width=\columnwidth]{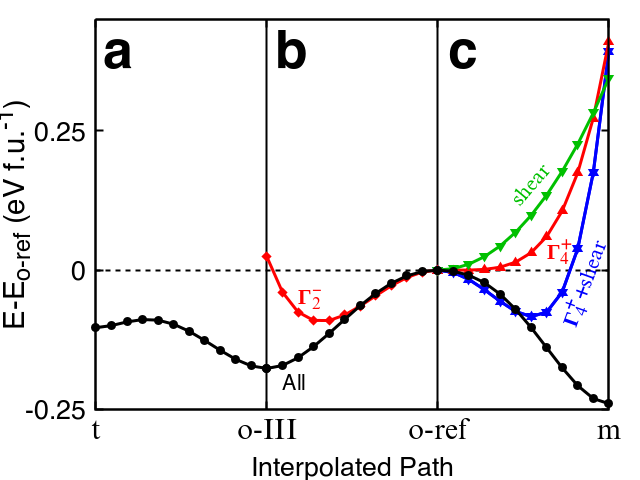}
    \caption{{\bf Energy landscape connecting key HfO$_{2}$
        polymorphs.} The black lines show the computed energy
      variation between the t and o-III phases ({\bf a}), o-III and
      o-ref ({\bf b}), and o-ref and m ({\bf c}). The energies are
      computed for intermediate structures obtained by linear
      interpolation between the corresponding end-point
      polymorphs. The red line in {\bf b} shows the energy variation
      of the o-ref state upon condensation of the $\Gamma_{2}^{-}$
      distortions present in the o-III phase; the red line in {\bf c}
      shows the analogous result when considering only the
      $\Gamma_{4}^{+}$ phonon distortions present in the m-phase. The
      blue line in {\bf c} shows the result of condensing together the
      $\Gamma_{4}^{+}$ phonon and shear strain distortions present in
      the m-phase, while the green line shows the energy variation
      associated to the shear alone. In {\bf b} and {\bf c}, the
      additional distortions leading to the black line are fully
      symmetric $\Gamma_{1}^{+}$ modes, including the normal cell
      strains.}
    \label{fig:paths}
\end{figure}

An additional issue pertains to the nature of
ferroelectricity. Figure~\ref{fig:paths}{\bf a} shows the energy of
hafnia computed along a path connecting t and o-III. (See Methods for
calculation details.) Both structures are local energy minima,
separated by a barrier, suggesting that a transition
  between them should be discontinuous (first-order).  Further, if
the energy landscape is analyzed in terms of symmetry-adapted
modes~\cite{delodovici21}, one concludes that the o-III state displays
an improper ferroelectric behavior, whereby the spontaneous
polarization relies on the occurrence of several non-polar modes. This
picture is appealing, because hafnia exhibits features typical of
improper ferroelectrics (e.g., large coercive fields,
 resilience of the polar order at the
  nanoscale). However, fresh experiments \cite{schroeder22} question
this interpretation.

Schroeder {\sl et al}. \cite{schroeder22} have recently reported a
strong dielectric anomaly ($\epsilon_{r}\approx 8000$) upon heating
Hf$_{0.5}$Zr$_{0.5}$O$_{2}$ across what seems to be a ferroelectric
phase transition. The authors note that the observed behavior is
reminiscent of proper ferroelectrics like BaTiO$_{3}$, which presents
a first-order transition driven by a soft mode with a high
permittivity maximum ($\epsilon_{r}\approx 10000$) associated to
it~\cite{jona-book1993}. In line with previous
  experimental \cite{park15,park18,tashiro21} and theoretical
  \cite{wu21,ganser22} reports of a transition between the t and o-III
  phases, the authors assume that the measured
  dielectric anomaly occurs at the temperature where the ferroelectric
  polymorph becomes unstable in favor of the tetragonal one. There is
a difficulty with this picture, though: it is not
  consistent with (in fact, it goes against) what we know about proper
  soft-mode ferroelectricity in perovskite oxides. Indeed, DFT
simulations of BaTiO$_{3}$ show that the paraelectric cubic state
presents a dominant polar instability~\cite{kingsmith94,ghosez99},
which is the hallmark of proper ferroelectricity driven by a soft
mode. The situation in hafnia is different: according to DFT, the
t-phase presents no indication of a polar instability (see
Supplementary Figure~S1) and, thus, there is no support for proper
soft-mode ferroelectricity. If anything, DFT suggests that a
transition between o-III and t would be similar to the
ferroelectric-paraelectric transition in BiFeO$_{3}$~\cite{catalan09}:
a non-soft-mode transformation between two states
that DFT describes as local energy minima~\cite{dieguez11}, and which
experimentally has a weak dielectric anomaly ($\epsilon_{r} \approx
65$) associated to it~\cite{polomska74,arnold09}.

Let us now discuss the picture of hafnia that emerges
  if we consider a different centrosymmetric reference, namely, the
orthorhombic $Pbcm$ state denoted ``o-ref'' in
Fig.~\ref{fig:domains}{\bf b}. This structure is similar to o-III,
except that the active oxygens (red in the figure) are located within
the same plane as the Hf atoms. This o-ref phase has been discussed in
theoretical investigations of ferroelectric switching as a potential
transition state~\cite{clima14,choe21,qi22}. Further,
  this polymorph has probably been observed experimentally in
  ZrO$_{2}$ under pressure \cite{kudoh86} and as an adaptive
  martensite phase in ZrO$_{2}$ nanoparticles~\cite{liu14}. At any
rate, let us stress that, for the present purposes, it is not critical
whether this polymorph actually occurs or not. Note that, for example,
the cubic paraelectric phase of BiFeO$_{3}$ is all but inaccessible
experimentally~\cite{arnold10}; yet, it is the relevant reference to
explain the observed ferroelectric domains.

\begin{figure*}
    \centering
    \includegraphics[width=\textwidth]{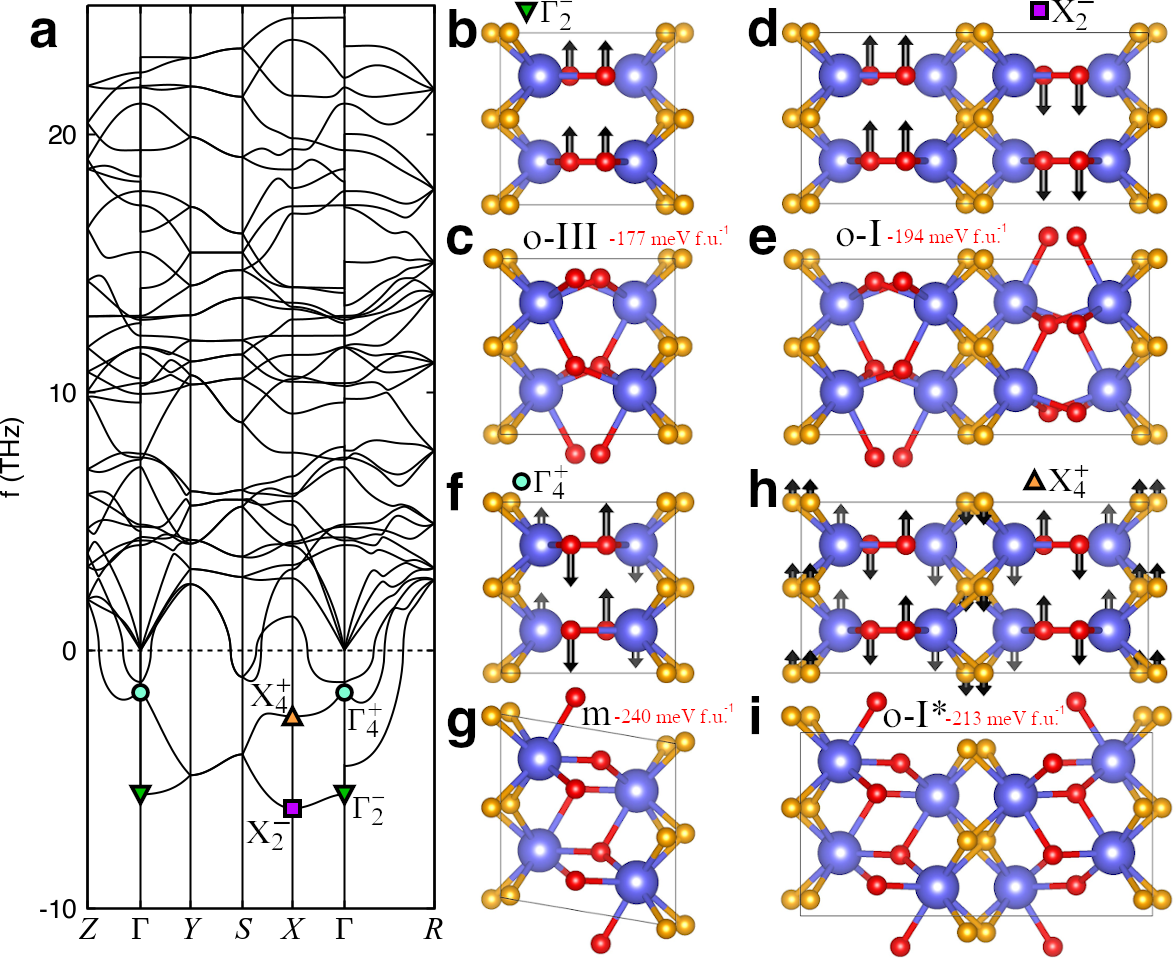}
    \caption{{\bf Phonon bands of the o-ref phase.} {\bf a} shows the
      computed bands, presenting imaginary frequencies as negative
      values. The most important unstable modes are marked in {\bf a}.
      We also show the corresponding eigenmodes and the polymorphs
      they lead to: the $\Gamma_{2}^{-}$ soft mode ({\bf b}) and the
      corresponding o-III phase ({\bf c}); the $X_{2}^{-}$ soft mode
      ({\bf d}) and the corresponding o-I phase ({\bf e}); the
      $\Gamma_{4}^{+}$ soft mode ({\bf f}) and the associated m-phase
      ({\bf g}); and the $X_{4}^{+}$ soft mode ({\bf h}) and the
      corresponding o-I* phase ({\bf i}). We mark in red the active
      oxygens whose displacements characterize these phonons. For the
      polymorphs, we indicate the energy with respect to o-ref.}
    \label{fig:phonons}
\end{figure*}

Figure~\ref{fig:phonons}{\bf a} shows the phonon bands computed for
the o-ref phase. We find a small number of unstable bands with
well-defined character: they feature off-plane displacements of the
active oxygens. At the $\Gamma$ point of the Brillouin zone, there is
a dominant instability: a polar mode with symmetry $\Gamma_{2}^{-}$
that captures the distortion connecting o-ref and o-III
(Figs.~\ref{fig:phonons}{\bf b} and~\ref{fig:phonons}{\bf c}). Indeed,
standard symmetry analysis~\cite{campbell06,aroyo06a} shows that o-III
can be obtained from o-ref by distortions of $\Gamma_{2}^{-}$ and
$\Gamma_{1}^{+}$ symmetries, the latter being fully-symmetric modes
not relevant in this discussion. The black curve in
Fig.~\ref{fig:paths}{\bf b} shows the energy variation associated to
this transformation: a simple potential well with a minimum about
177~meV per formula unit (f.u.) below o-ref. Thus, an
  energy expansion around o-ref naturally provides a framework to
  describe soft-mode-driven proper ferroelectric behavior, as that of
  BaTiO$_{3}$. Further, the $\Gamma_{2}^{-}$ irreducible
representation is one dimensional; thus, we have uniaxial
ferroelectricity, as sketched in Fig.~\ref{fig:domains}{\bf b}.

Incidentally, let us note that, as discussed in Supplementary Note~1
and previous literature~\cite{choe21,qi22}, using o-ref instead of the
t-phase as a centrosymmetric reference leads to a different -- but
related within the Berry-phase theory~\cite{kingsmith93} --
quantification of the polarization of the o-III phase.

If we follow the $\Gamma_{2}^{-}$ soft mode along the $\Gamma-X$
direction of the Brillouin zone, we reach a stronger instability: the
antipolar $X_{2}^{-}$ mode in Fig.~\ref{fig:phonons}{\bf d}. This
distortion yields another low-energy polymorph: the orthorhombic
$Pbca$ phase depicted in Fig.~\ref{fig:phonons}{\bf e}, which is
usually denoted ``o-I'' \cite{azevedoantunes22} and has been
experimentally observed in ZrO$_{2}$~\cite{ohtaka90}. Symmetry
analysis shows that o-ref and o-I are connected solely by $X_{2}^{-}$
and fully-symmetric distortions. The o-I state is more favorable than
the ferroelectric o-III phase, by about 17~meV per f.u. Further, this
state can be seen as an sequence of ultra-thin ferroelectric stripe
domains, where the polarization of the o-III phase is modulated along
the horizontal direction in Fig.~\ref{fig:phonons}{\bf b}. From this
perspective, we can say that the domain walls of the o-III phase have
a negative formation energy (of about $-84$~mJ~m$^{-2}$) and that
ferroelectricity in hafnia is essentially two-dimensional (i.e., a
single ultra-thin ferroelectric stripe can occur regardless of its
surroundings). This and related observations (e.g., the slow motion of
domain walls in hafnia) had been previously made~\cite{lee20,qi21};
here, by taking the o-ref phase as a starting point, we rediscover
them in a straightforward manner.

Our calculations reveal additional soft modes worth
discussing. Especially interesting is the second lowest-lying
zone-center instability, with $\Gamma_{4}^{+}$ symmetry and an
antipolar character (Fig.~\ref{fig:phonons}{\bf f}). This distortion
leads to the well-known monoclinic ground state of hafnia ($P2_{1}/c$
``m-phase'', Fig.~\ref{fig:phonons}{\bf g}), about 240~meV per
f.u. below o-ref~\cite{azevedoantunes22,barabash17}. If we follow the
$\Gamma_{4}^{+}$ instability as we move towards $X$, we reach a soft
mode with symmetry $X_{4}^{+}$. Sketched in Fig.~\ref{fig:phonons}{\bf
  h}, this mode involves an antiphase modulation of the antipolar
distortion in Fig.~\ref{fig:phonons}{\bf f}; its condensation yields
the state shown in Fig.~\ref{fig:phonons}{\bf i}, with space group
$Pbca$ and about 213~meV per f.u. below o-ref, denoted
``o-I*''~\cite{azevedoantunes22} and experimentally
observed~\cite{ohtaka95,du21}. Note that the m and o-I* phases are
both connected to the o-ref state by distortions of well-defined
symmetry ($\Gamma_{4}^{+}$ and $X_{4}^{+}$, respectively) solely
accompanied by fully-symmetric modes. Hence, notably, the most
relevant low-energy polymorphs of hafnia (m, o-III, o-I and o-I*) can
be obtained as simple proper instabilities of the o-ref phase.
Additional low-energy structures -- e.g., the ``m-III'' polar phase
recently discussed in Ref.~\onlinecite{azevedoantunes22} -- can also
be obtained, by condensing other individual soft modes or combinations
of them. Supplementary Figure~S2 shows that essentially the same
applies to ZrO$_{2}$.

Note that our proposed ``uniaxial approach'' to hafnia
  affects not only the o-III phase, but also the mentioned low-lying
  polymorphs. All of them share an axis along which the active oxygens
  move, namely, the vertical direction that is common to all the
  structures of Fig.~\ref{fig:phonons}. For convenience, Supplementary
  Figure S4 shows the simulated X-ray diffraction patterns for all
  these phases, and Supplementary Table~S1 lists the corresponding
  lattice constants as obtained from our simulations.

The phonon frequencies of Fig.~\ref{fig:phonons}a pose
  an apparent paradox: given that the computed $\Gamma_{4}^{+}$
instability is weak compared to the others discussed above, how can
the associated m-phase be the ground state? To answer this, in
Figs.~\ref{fig:paths}{\bf b} and \ref{fig:paths}{\bf c} we distinguish
the energy contributions of different sets of modes to stabilize the
o-III and m phases. The condensation of the $\Gamma_{2}^{-}$
distortion yields a large energy reduction with respect to o-ref (red
curve in Fig.~\ref{fig:paths}{\bf b}); then, the fully-symmetric
$\Gamma_{1}^{+}$ modes react to the ferroelectric distortion and
further reduce the energy (black curve) down to the actual o-III
minimum. By contrast, the condensation of the $\Gamma_{4}^{+}$ optical
distortion alone (red curve in Fig.~\ref{fig:paths}{\bf c}) yields a
shallow energy well. Nevertheless, in the case of the m-phase, the
shear strain causing the monoclinic deformation of the cell shares the
$\Gamma_{4}^{+}$ symmetry. Hence, while stable by itself (green
curve), this shear couples harmonically with the $\Gamma_{4}^{+}$
phonon yielding a much stronger instability (blue curve). In addition,
the $\Gamma_{1}^{+}$ modes react to the monoclinic distortion,
resulting in the very stable m-phase (black curve). Hence, by using
the o-ref phase as the starting point of our analysis, we reveal the
key role of the shear strain in determining the ground state of
hafnia, reflected in the energetics of Fig.~\ref{fig:paths}{\bf c} and
the fact that the total $\Gamma_{4}^{+}$ instability has a mixed
strain-phonon character.

Our approach also sheds light into the possible transitions between
stable hafnia polymorphs. For example, an electric field-driven
transformation from o-I to o-III would constitute a textbook case of
antiferroelectric behavior. Indeed, because the antipolar
($X_{2}^{-}$) and polar ($\Gamma_{2}^{-}$) instabilities belong to the
same band (Fig.~\ref{fig:phonons}{\bf a}), this appears to be an ideal
``Kittel antiferroelectric''~\cite{kittel51}. Further, we suggest that
field-driven transformations from o-I* to o-III, or from m to o-III,
can also be viewed as Kittel-like antiferroelectric transitions, since
the polar and antipolar states have a common origin as distortions of
the o-ref phase. The latter (m to o-III) would be a rare example of
antiferroelectric effect involving no doubling of the unit
cell. Along these lines, let us note that Kudoh {\sl
    et al}. \cite{kudoh86} studied ZrO$_{2}$ under pressure and found
  what seems to be an order-disorder transition from m to o-ref
  itself, with oxygens hopping back and forth across their
  high-symmetry positions in the $Pbcm$ structure.

\begin{figure}
    \centering
    \includegraphics[width=\columnwidth]{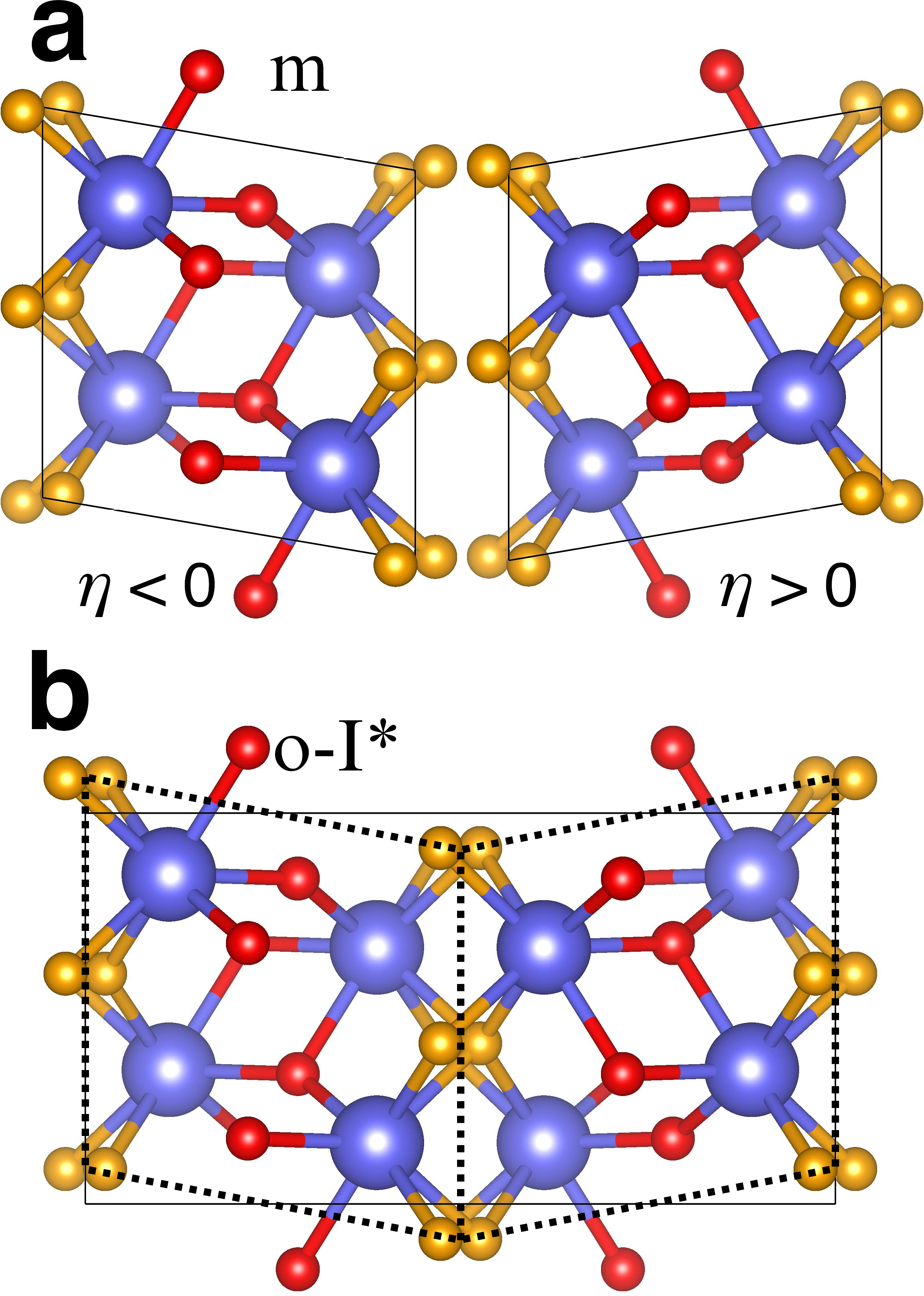}
    \caption{{\bf Antiferroelastic behavior.} {\bf a} shows the two
      symmetry-equivalent m-phase variants that can be obtained as a
      $\Gamma_{4}^{+}$ distortion of the o-ref phase. In {\bf b} we
      emphasize that the unit cell of the o-I* phase can be obtained
      by matching the two m-variants in {\bf a}. Hence, o-I* can be
      seen as composed of ultra-thin ferroelastic stripe domains, with
      a domain wall energy of 126~mJ~m$^{-2}$.}
    \label{fig:anti}
\end{figure}

Our results also suggest that o-I* is structurally connected to the
monoclinic ground state. In essence, this connection was already
mentioned by Ohtaka {\sl et al}~\cite{ohtaka95}; our theory confirms
it and reveals its deeper origin. Indeed, viewed as a distortion of
o-ref, the m-phase can present the two variants shown in
Fig.~\ref{fig:anti}{\bf a}, with positive and negative shear,
respectively. Then, as emphasized in Fig.~\ref{fig:anti}{\bf b}, the o-I*
unit cell can be seen as composed of two matching domains
corresponding to such m-variants. This is most natural: the underlying
$\Gamma_{4}^{+}$ and $X_{4}^{+}$ instabilities belong to the same
band, which suggests that the associated distortions correspond to
different modulations of the same local motif. Such common motif is
the particular antipolar displacement of the active oxygens in
Fig.~\ref{fig:phonons}{\bf f}, which is accompanied by a relative vertical
shift of the neighboring Hf planes. When this pattern repeats
homogeneously ($\Gamma_{4}^{+}$) it yields a net shear; when
antimodulated ($X_{4}^{+}$) the opposing local strains cancel
out. These observations further suggest that a transition from o-I* to
m, driven by an appropriate shear stress, would be an example of
``antiferroelastic'' behavior. Interestingly, antiferroelastics were
introduced theoretically decades ago~\cite{aizu69}, but we are not
aware of any demonstration. (The term ``antiferroelastic'' has been
used to denote phases that present correlated antiferrodistortive and
Jahn-Teller distortions~\cite{watanabe06}; however, as far as we can
see, such phases do not involve the antiphase strain modulations
discussed by Aizu~\cite{aizu69} and shown here.) In the case of
hafnia, an experimental realization would require stabilizing the o-I*
state over the m-phase; given their proximity in energy (the gap is
about 27~meV per f.u.) this is conceivable, e.g., by growing samples
on appropriate substrates.

We now make some remarks that put our work into perspective. Let us
start by noting that Zhou {\sl et al}.~\cite{zhou22b} have recently
used DFT to predict that, under suitable elastic constraints, the
t-phase evolves into an antipolar state with $Pbcn$ symmetry, which in
turn can eventually develop a ferroelectric soft mode leading to
o-III. Raeliarijaona and Cohen \cite{raeliarijaona23}
  have provided an additional DFT-based discussion on how the
  centrosymmetric $Pbcn$ state can yield proper ferroelectric
  order. We think these theories may indeed prove relevant to explain
  the experimentally observed dielectric anomaly and soft-mode-like
  behavior~\cite{schroeder22}. Interestingly, both these works imply
  an uniaxial picture of hafnia, which is justified as a result of
  epitaxial strain by Zhou {\sl et al}.~\cite{zhou22b}.

On the topic of the temperature-driven transition, it
  is worth noting that Ref.~\onlinecite{liu14} showed experimental
  evidence -- for zirconia nanoparticles -- of a martensitic
  transformation from t to o-I through an intermediate o-ref
  state. Given the structural similarities between the o-I and o-III
  states (see Figs.~\ref{fig:phonons}c and \ref{fig:phonons}e), these
  results suggest that a transition between o-III and t may present an
  intermediate orthorhombic centrosymmetric state like o-ref. If so,
  this could explain the soft-mode-like dielectric anomaly observed
  experimentally upon heating the o-III state~\cite{schroeder22}.

Let us briefly mention that machine-learned potentials
  -- derived from DFT simulations -- have been used to study the
  temperature-driven ferroelectric-to-paraelectric phase transition in
  both hafnia \cite{wu21} and zirconia~\cite{ganser22}. Both works
  report an o-III to t transformation, involving a dielectric anomaly
  in the case of zirconia ($\epsilon_{r}\approx 330$ at
  900~K)~\cite{ganser22}. While valuable, we are not convinced these
  simulations explain the experimentally observed
  transformation~\cite{schroeder22}. On the one hand, while it
  reflects some lattice softening, the observed dielectric maximum is
  quite modest compared to the experimental one observed in hafnia
  ($\epsilon_{r}\approx 8000$). On the other hand, the results of
  these studies probably depend on the training sets used to construct
  the machine-learned potentials, and we wonder whether the
  orthorhombic states discussed here or in Refs.~\onlinecite{zhou22b}
  and \onlinecite{raeliarijaona23} were considered. Hence, we think
  this remains an open question.

Let us now consider ferroelectric switching or
  field-driven transitions between the low-lying polymorphs. As
  already mentioned, the latest experimental \cite{li23} and
  theoretical \cite{silva23} results suggest that the most favorable
  switching paths involve hopping of the active oxygens across their
  high-symmetry positions in the o-ref state, giving rise to local
  configurations that resemble the structural motifs of the m, o-I and
  o-I* polymorphs. Further, these low-energy paths do not seem to
  involve t- or $Pbcn$-like configurations.
 Interestingly, the situation is reminiscent of the
  order-disorder hopping of active oxygens, around their equilibrium
  positions in the o-ref phase, as deduced from experiments of
  ZrO$_{2}$ under pressure~\cite{kudoh86}. Such a dynamical disorder
  can be viewed as (local) stochastic transitions between the m, o-I,
  o-I* and o-III states, yielding the o-ref structure in average.
 Hence, in order to investigate ferroelectric switching
  and field-driven transitions in woken-up hafnia and zirconia, it
  seems most suitable to construct a theory that takes the o-ref state
  as starting point, so that all relevant intermediate states can be
  described as simple distortions of the reference structure.

We thus have the following remarkable situation. For
  hafnia and zirconia, we may need to work with a tetragonal or cubic
  reference to study the stabilization of the o-III ferroelectric
  state, including wake-up cycling. This implies a biaxial or triaxial
  material, respectively.  By contrast, it may be sufficient, and
  physically more transparent, to build uniaxial theories based on the
  $Pbcn$ or o-ref structures in order to discuss the proper
  ferroelectric phase transition and soft-mode-like dielectric anomaly
  observed experimentally. Finally, to investigate ferroelectric
  switching -- or field-driven transformations among low-lying
  polymorphs -- the o-ref state probably offers the simplest and
  physically most relevant starting point.

The above disquisition may seem puzzling, particularly
  if one has in mind the simplicity of the best studied ferroelectrics
  family, i.e., perovskite oxides like BaTiO$_{3}$ or
  PbTiO$_{3}$. There, the ideal (cubic) perovskite structure is
  generally taken as the high-symmetry centrosymmetric reference, for
  all materials and purposes. However, even among perovskites there
  are subtle cases that can inform our present discussion. For
  example, the ferroelectric transition of BiFeO$_{3}$ involves a
  complex paraelectric phase featuring large distortions of the ideal
  cubic structure (large tilts of the O$_{6}$ octahedra, $Pbnm$ space
  group). This non-polar state bears no group-subgroup symmetry
  relation with the ferroelectric phase ($R3c$ space group); in fact,
  it would be hopeless to (try to) use the $Pbnm$ state as the
  starting point of a theory aiming to explain the properties (domain
  variants, switching) of the ferroelectric phase. Hence, BiFeO$_{3}$
  is an example where the theory needed to explain the ferroelectric
  phase transition (which must account for the $Pbnm$ phase) is
  different from the theory needed to model the properties of the
  ferroelectric state (which requires a cubic reference).

It is also worth noting the case of
  LiNbO$_{3}$~\cite{weis85}, which features a rhombohedral ($R3c$)
  ferroelectric phase that is very similar to that of BiFeO$_{3}$. By
  analogy with BiFeO$_{3}$, one would postulate a cubic reference and
  treat this compound as a triaxial ferroelectric. However,
  experiments show that LiNbO$_{3}$ behaves as a uniaxial material,
  and that a centrosymmetric rhombohedral ($R\bar{3}c$) phase is the
  paraelectric state~\cite{scrymgeour05}. Hence, a simple model taking
  the $R\bar{3}c$ phase as reference -- and yielding only two
  polarization variants -- gives a description of LiNbO$_{3}$ that is
  both simple and sufficient. LiNbO$_{3}$ thus provides us with an
  example where phenomenological arguments -- not unlike the ones made
  in this work -- lead us to a theory that is simpler and physically
  sounder than the one that might have been chosen by default. (See
  Supplementary Note~2 for further considerations on the choice of
  reference states in BiFeO$_{3}$ and LiNbO$_{3}$.)

In summary, in this work we introduce a theoretical
  framework ideally suited to model the functional properties of the
  most common ferroelectric phase of hafnia and zirconia, including
  switching, field-driven transitions between low-energy polymorphs,
  and electromechanical responses. We focus on the description of the
  woken-up ferroelectric state, and rely on the -- experimentally
  supported -- assumption of a uniaxial ferroic order. We also discuss
  how other phenomena may require different treatments, potentially
  involving alternative reference phases (e.g., to address the
  temperature-driven ferroelectric transition) and abandoning the
  uniaxial hypothesis (e.g., to tackle ferroelastic effects during the
  wake-up process).

Within this restricted -- but crucial -- realm of application, our
work provides a simple yet thorough picture of the relevant energy
landscape of hafnia and zirconia, naturally connecting all low-energy
polymorphs. In particular, the proposed reference phase appears as an
ideal starting point for the development of physically transparent
perturbative theories, from phenomenological Ginzburg-Landau models to
coarse-grained effective Hamiltonians~\cite{zhong94a} or atomistic
second-principles potentials~\cite{wojdel13}. Further, having such a
soft-mode-style model of hafnia invites (enables) us to borrow ideas
from the literature on perovskite oxides, for example, to optimize the
negative capacitance effect~\cite{iniguez19,graf22}. We thus expect
our findings will become an important ingredient of future work on
these materials, from theoretical and computational studies to the
conception of new experiments and optimization strategies.

{\bf Methods.} Our simulations are carried out using first-principles
density functional theory (DFT) as implemented in the Vienna Ab-initio
Simulation Package (\textsc{VASP})~\cite{kresse96,kresse99}.  We
employ the Perdew-Burke-Ernzerhof formulation for solids
(PBEsol)~\cite{perdew08} of the generalized gradient approximation for
the exchange-correlation functional.  The atomic cores are treated
within the projector-augmented wave approach~\cite{blochl94},
considering the following states explicitly: 5s, 5p, 6s, 5d for Hf;
4s, 4p, 5s, 4d for Zr; and 2s, 2p for O. We use a plane-wave energy
cutoff of 600~eV. A 6$\times$6$\times$6
Monkhorst-Pack~\cite{monkhorst76} $k$-point sampling of the Brillouin
zone is employed for the o-ref, m, t and o-III phases, and a
3$\times$6$\times$6 $k$-point grid is employed for the o-I and o-I\*
phases, which are (approximately) twice as long along the first
lattice vector. The structures are fully relaxed until the residual
forces fall below 0.001~eV \AA$^{-1}$ and residual stresses fall below
0.01~GPa. These calculation conditions yield well-converged results.

The paths shown in Fig.~\ref{fig:paths} are obtained by interpolating
the lattice vectors and fractional atomic coordinates between the
initial and final structures with 10 intermediate points. No
structural optimization is performed for the intermediate
states. Thus, for example, the green points in the
  figure are obtained by computing the energy of the o-ref structure
  distorted by a shear strain as the one of the m-phase, from zero
  shear (pristine o-ref) to its value in the m polymorph. The blue
  points give the energy of an o-ref structure distorted by condensing
  simultaneously the shear strain and the $\Gamma_{4}^{+}$ phonon
  distortion that appear in the m-phase, from zero (pristine o-ref) to
  their values in the m polymorph.

The polarization is computed using the modern theory of
polarization~\cite{kingsmith93}. Phonon bands are obtained using the
direct supercell approach implemented in the \textsc{phonopy}
package~\cite{phonopy}. A 2$\times$2$\times$2 supercell is employed
both for the o-ref phase of HfO${_2}$ and ZrO${_2}$ and for the
t-phase of HfO${_2}$, which we find to give well-converged results.
The non-analytical contribution to the phonons is considered in the
calculations.

We use standard web-based crystallographic tools
\cite{campbell06,aroyo06a} for symmetry analysis. The visualization
package \textsc{vesta}~\cite{vesta11} is used for the structure
representations and to simulate X-ray diffraction
  patterns.

{\bf Acknowledgments}. Work supported by the Luxembourg National
Research Fund through Grant INTER/NWO/20/15079143/TRICOLOR.

\end{document}